\documentclass[
twocolumn, showpacs, prl,
 superscriptaddress,
 ]{revtex4}%

\usepackage{color}

\newcommand{\blambda}{
\mbox{$\lambda \hspace*{-0.49em} \protect\raisebox{0.32em} {$
\mathbf {\scriptstyle
    -}$}\hspace*{-0.07em}
$} }

\newcommand{\nnabla}{\boldsymbol{\nabla}}

\newcommand{\req}[1]{Eq.~(\ref{#1})}
\newcommand{\reqs}[1]{Eqs.~(\ref{#1})}
\newcommand{\rref}[1]{(\ref{#1})}

\newcommand{\p}{\mathbf{p}}
\newcommand{\n}{\mathbf{n}}
\renewcommand{\r}{\mathbf{r}}
\newcommand{\R}{\mathbf{R}}
\renewcommand{\k}{\mathbf{k}}

\newcommand{\beq}{\begin{equation}}
\newcommand{\eeq}{\end{equation}}
\newcommand{\be}{\begin{equation}}
\newcommand{\ee}{\end{equation}}
\newcommand{\beqa}{\begin{eqnarray}}
\newcommand{\eeqa}{\end{eqnarray}}
\newcommand{\bea}{\begin{eqnarray}}
\newcommand{\eea}{\end{eqnarray}}
\usepackage{amssymb}
\usepackage{array}
\usepackage{amsmath}
\usepackage{graphicx}
\usepackage{dcolumn}
\usepackage{bm}
\usepackage[mathscr]{eucal}

\usepackage{amsfonts}%

\usepackage{bm}
\usepackage{amsmath}
\usepackage{amssymb}
\usepackage{graphicx}

\begin{document}

\title{Effect of Disorder on Transport in Graphene}

\author{I.~L.~Aleiner}
\affiliation{Physics Department, Columbia University, New York, NY
10027, USA}

\author{K.~B.~Efetov}

\affiliation{Physics Department, Columbia University, New York, NY
10027, USA}
\affiliation{Theoretische Physik III, Ruhr-Universit\"at Bochum,
44780 Bochum, Germany } \affiliation{L. D. Landau Institute for
Theoretical Physics, 117940 Moscow, Russia}
\date{\today}

\begin{abstract}
Quenched disorder in graphene is characterized by 5 constants and
experiences the logarithmic renormalization even from the spatial
scales smaller than the Fermi wavelength. We derive and solve
renormalization group  equations (RGEs) describing the system at
such scales. At larger scales, we derive a non-linear supermatrix
$\sigma $-model completely describing localization  and
crossovers between different ensembles. The parameters of this $\sigma $
-model are determined by the solutions of the RGEs.
\end{abstract}

\maketitle

{\em Introduction}--
Dirac spectrum of the quasiparticles in graphene confirmed by recent experiments
\cite{novo,zhang,berger} is a consequence of the honeycomb lattice
symmetry \cite{Pikus}. Although many properties of graphene can be
understood in terms of ballistic motion of \textquotedblleft
relativistic\textquotedblright\ electrons described by a Dirac-like
equation (see, e.g. \cite{gusynin,Kats}), disorder  plays an important role in
sufficiently large samples.

Influence of disorder on a two-dimensional
 electron gases on the honeycomb lattice were studied in
several works \cite{shon,suzuura,peres,zheng,ando,Morpurgo,falko}
 within a self-consistent Born
approximation (SCBA) standard  for weakly disordered metals and superconductors
\cite{agd}. Within SCBA, such quantities as density of states,
or localization-less
conductivity \cite{suzuura,shon,zheng,ando}
were calculated.
The  weak localization (WL) correction was discussed in
Ref.~\cite{Morpurgo} and calculated
in Ref.~\cite{falko}.

However,
the SCBA is not justifiable for  the Dirac
spectrum, $|\varepsilon|=v|p|$, as it can be seen already in the fourth
order perturbation theory in the disorder potential, see below,
and a more careful analysis is needed.

\begin{figure}[h]
\unitlength=2.3em
{\includegraphics[width=9.1\unitlength]{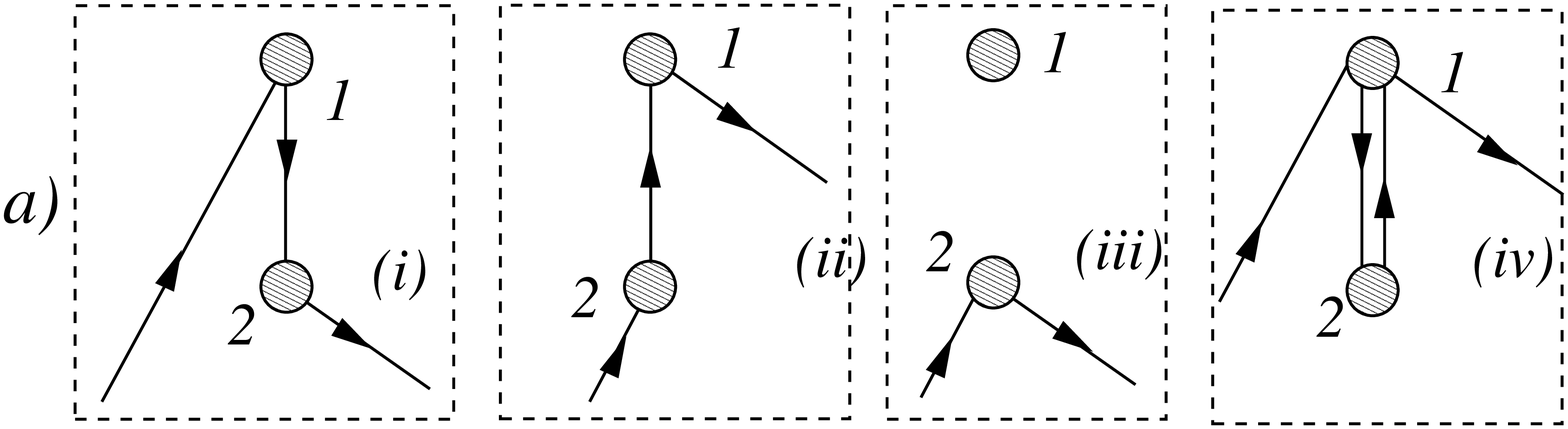}}
\\
{\includegraphics[width=10.5\unitlength]{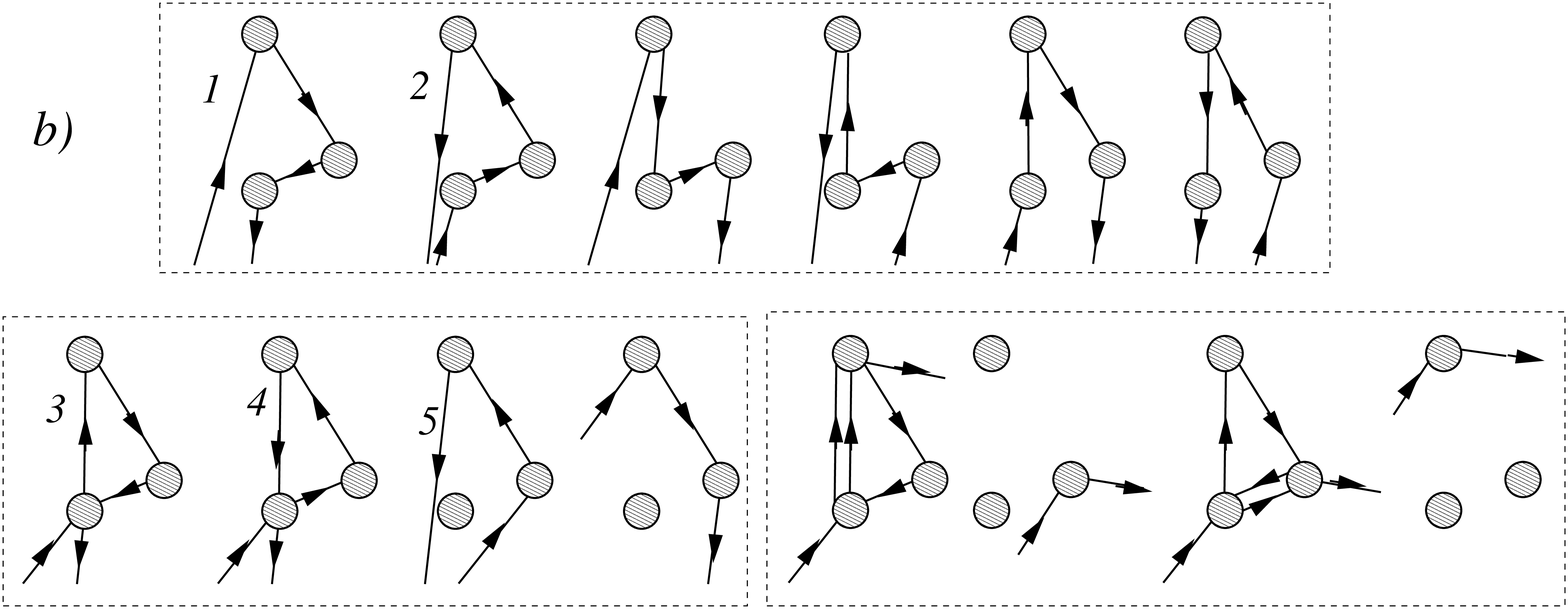}}

\caption{Interfering scattering events involving
(a) two- and (b) three-impurity scattering. They produce the correction
(a) ${\mathcal O} \left[ \ln\left(\protect{\blambda}_\varepsilon/a_0
\right) \right] $ and (b) ${\mathcal O} \left[ \ln^2\left(\protect{\blambda}_\varepsilon/a_0
\right) \right] $.}

\label{fig1}
\end{figure}

In this paper, we reveal the  origin of the logarithmic effects
specific for the Dirac spectrum and different from WL. These
corrections, see Fig.~\ref{fig1}, are contributed by all spatial scales between the
lattice constant $a$, and either wave-length
$\blambda_\varepsilon=\hbar v/\varepsilon$, or the scale
determined by disorder, and that is why we will coin the name of
``ultraviolet logarithmic corrections'' (UvLC) for them. We will
sum up the leading series of UvLC within a one-loop RG. At
larger linear scales all the physics is described by a non-linear
$\sigma $-model\cite{book} and UvLC enter as renormalized
parameters in this model.  We will
show that the  low-energy asymptotics  correspond
to the orthogonal ensemble. Thus, the one-particle states are
localized at any energies in contradiction with the findings of
Ref.~\cite{novo} of the minimal metallic conductivity
in the undoped graphene.

{\em Disordered Hamiltonian}--
In the undoped graphene two bands cross the Fermi level at $K$ and $%
K^{\prime }$ points. The corresponding Bloch functions comprise
the basis of the four-dimensional (4d) representation of the
planar group of the honeycomb lattice. We  join them in a vector
\be
\vec{\varphi}^T(\r)=\left(\left(\varphi_A,\varphi_B\right)_{AB};
\left(\varphi^*_B,-\varphi^*_A\right)_{AB} \right)_{KK'}
\label{bloch} \ee where we use the fact the points $K$ and $K'$
are connected to each other by the time reversal symmetry \be
\vec{\varphi}^*(\r)=\hat{z}\vec{\varphi}(\r); \quad
\hat{z}\equiv\tau_y^{AB}\otimes\tau_y^{KK'}. \label{TR} \ee where
4d-space of the wave-functions is represented as a
direct product $AB\otimes KK'$, of the sublattice $AB$ and
``valley'' $KK'$ 2d-spaces, and  $\tau_{x,y,z}^\alpha$,
$\tau^\alpha_\pm=(\tau_x\pm i\tau_y)/2$ are the Pauli matrices
acting in those spaces, $\alpha=AB,\ KK'$ (we omit the physical
spin).

The low energy properties are described in
$k\cdot\p$-approximation \cite{Pikus}, i.e. the wave function
$\Psi_\varepsilon(\r)$ is looked for as
$\Psi_\varepsilon(\r)=\vec{\varphi}(\r)
\cdot\vec{\Phi}_\varepsilon(\r)$ with a smooth envelope,
$\vec{\Phi}_\varepsilon(\r)$,  satisfying the effective
Schr\"{o}dinger equation ($\hbar=1$)
\begin{equation}
\left( \hat{H}_{0}+\hat{V}(\r)\right)\vec{ \Phi}_\varepsilon(\r) =\varepsilon\vec{\Phi}_\varepsilon(\r),
\quad \hat{H}_{0}=-iv\vec{\hat\Sigma}{\vec{\nabla}},
\label{a1}
\end{equation}
where $\vec{\nabla}=(\partial_x,\partial_y)$ and we introduced $4\times 4$ matrices
\cite{footnote1}
\be
\hat{\Sigma}_{x,y}=\openone^{KK'}\otimes \tau_{x,y}^{AB}, \quad
\hat{G}_{m,i}
= \tau_{m}^{KK'}\otimes\tau_{i}^{AB}.
\label{a2}
\ee

For the spinless particles, the  time-reversal symmetry (TRS)
requires $\Psi^*(\r)=\vec{\varphi}^*(\r)\cdot\vec{\Phi}^*(\r)$
also to be the eigenstate of the original Hamitltonian. Together
with \req{TR} it constrains  the effective Hamiltonian from
\req{a1} as
\begin{equation}
\hat{H}_0=\hat{z}\hat{H}^{T}_0\hat{z}, \quad
\hat{V}=\hat{z}\hat{V}^{T}\hat{z}.  \label{a5}
\end{equation}%

The disorder, $V(\r)=V^{\dagger}(\r)$, in \req{a1} 
may break all the symmetries except of the TRS
\rref{a5}. This leads to 
\begin{equation}
\hat{V}(\r)=\openone u_{0}(\r)+\hat{G}_{m,i}u_{m,i}(\r),  \label{a6}
\end{equation}%
where  $u_{0}$, $u_{m,i}$ are real random functions, $\openone\equiv\openone^{AB}\otimes \openone^{KK'}$,
and the summation over indices $\{i,m\}=\{x,y,z\}$ is implied
in \reqs{a6}--\rref{a8}.
Equation \rref{a6} is the most general form for the
disorder \textit{before} averaging. Averaging
must restore the rotation, reflection, ($C_{6v}$), and translation symmetries,
which leads\cite{ae} to the most general Gaussian correlations
($\gamma_0>0,\Gamma _{m,i}>0$)
\begin{subequations}
\begin{equation}
\langle \hat{V}_\r\otimes \hat{V}_\r^\prime
\rangle
=\delta_{\mathbb{\r},\mathbb{\r'}}
\left[\gamma_0 \openone \otimes \openone+
\Gamma _{m}^{i}\hat{G}_{m,i}\otimes\hat{G}_{m,i}
\right].
 \label{a8}
\end{equation}%
Among the coefficients $\Gamma _{m,i}$ only four are
 independent:
\be
\label{a10}
\Gamma _z^z=\gamma _{z},\
\Gamma _z^{\{x,y\}}=\gamma _{\perp }, \ \Gamma _{\{x,y\}}^{z}=\beta _{z},\
\Gamma _{\{x,y\}}^{\{x,y\}} =\beta _{\perp }.
\ee
\label{a8a10}
\end{subequations}

In what follows, we will treat the five independent parameters
$\gamma,\beta$ of \reqs{a8a10} as the starting point of the
theory, not trying to calculate their particular values determined
by the details of the single impurity on the scale of the lattice
constant. However, some qualitative conclusions can be drawn from
the symmetry arguments. The diagonal term,
$(\gamma_0>0,\Gamma_{m}^i=0)$, is generated by the dilute
impurities such that a single impurity in the crystal still
preserves the point symmetry $C_{6v}$.  Examples are the
interstitial atom, or,
 more importantly, the remote charge impurity potential
weakly varying on the scale of the lattice constant, $a$. On the
other hand, a vacancy in otherwise perfect honeycomb lattice
preserves only $C_{3v}$ symmetry and allows for
$\gamma_z\simeq\gamma_0$. The intravalley scattering,
$\gamma_\perp>0$ arises e.g. due to the local deformation
\cite{iordanskii} or any bond disorder breaking the $C_3$
symmetry. The latter also causes the intervalley scattering,
$\beta_{z,\perp}$. Finally, the terms of the similar structure are
generated even by the multiple scattering by the pair of vacancies
if the higher gradient terms are included in Hamiltonian
\rref{a1}.

%\newpage
%The real coefficients $\gamma _{0\text{,}}\gamma _{z}$, $\gamma _{\perp }$
%in Eq. (\ref{a10}) relate to the intravalley disorder correlations, while $%
%\beta _{z}$ and $\beta _{\perp }$ stand for the intervalley scattering. The
%diagonal disorder with the correlation $\gamma _{0}$ originates from long
%range scatterers like charge impurities on the substrate and is the
%strongest one for the graphene systems. The intervalley scattering is due to
%the presence of sharp impurities and non-zero values of $\gamma _{z}$ and $%
%\gamma _{\perp }$ are possible only in this case. The correlations with $%
%\gamma _{z}$, $\beta _{z}$, $\gamma _{\perp },$ $\beta _{\perp }$ originate
%from bond defects and different potentials on the neighboring sites
%belonging to the different sublattices.
%\newpage
%Eqs. (\ref{a1}-\ref{a3}, \ref{a8}, \ref{a10}) describe completely the
%disordered graphene system and allow one to calculate transport properties.
%As in the theory of conventional disordered metals it is convenient to write
%physical quantities in terms of a functional integral over supervectors and
%average over disorder from the beginning \cite{book}. Then, one would find a
%saddle-point and derive a non-linear $\sigma $-model describing the
%localization.
%However, as we have mentioned, the SCBA and, hence, the saddle
%point approximation are generally not applicable. Before we
%proceed with formal computations, let us explain the physical
%reason for the failure of the SCBA.
{\em Origin of UvLC} lies in the interference of the waves
multiply scattered by the disorder potential, however, the
configurations of the impurities giving rise to the effect  differ
from the familiar \cite{book} WL. Thus, it is instructive to
explain them, first, in terms of the counting of the scattering
events and then present the rigorous calculation. It will also
illuminate reasons for the failure of SCBA.

If there were no disorder potential the wave function
$\vec{\Phi}^{(0)}_\k$ would be the plane-wave with with a momentum
$\k$ and the structure of \req{bloch}. Limit ourselves by positive
energies $\varepsilon= vk,\ k\equiv |\k|$ for concreteness, and
the projection operator $\hat{\mathbb P}(\n)\equiv
\left(1+\boldsymbol{\Sigma}\n\right)/2; \ \n=\k/|k| $ chooses the
chirality corresponding to the positive energy. Consider an
impurity  placed at point $\R_{1}$. The asymptotics of wave
function  acquire the form
$\vec{\Phi}_\k=\vec{\Phi}^{(0)}_\k+\vec{\Phi}^{(1)}_\k$, where \be
\begin{split}
\vec{\Phi}^{(1)}_\k\!\!=  \left\{
\begin{matrix}
\displaystyle{
\frac{\exp\left(\frac{i|\r_1|}{\blambda_\varepsilon}\right)}{\sqrt{-i|\r_1|}}
\frac{\hat{\mathbb P}\left(\frac{\r_1}{|\r_1|}\right)
\hat{f}_1}{\sqrt{2\pi\blambda_\varepsilon}} \vec{\Phi}^{(0)}_{\k}(\R_1) ;}
& \frac{|r_1|}{\blambda_\varepsilon}\gg 1;
\\
\displaystyle{ {i\left({2\pi|\r_1|^2}\right)^{-1}\hat{\boldsymbol{\Sigma}}\r_1}
\hat{f}_1    \vec{\Phi}^{(0)}_{\k}(\R_1)}
;&{|r_1|}\ll {\blambda_\varepsilon};
\end{matrix}
\right.
\end{split}
\raisetag{2em}{\label{eq:scat1}} \ee
and $\r_1\equiv\r-\R_1$ is
the distance from the impurity.

Equation \rref{eq:scat1} is valid for both $|\r_1|$ and
$\blambda_\varepsilon$ much larger than the characteristic size
of the scatterer. All the the details of
the impurity are encoded in the matrix $\hat{f}_1$
which
 can be viewed as the scattering length,
$a_0=\parallel\hat{f}_1\hat{f}_1^\dagger\parallel^{1/2}$. The
 value of $a_0$ is of the order of the size of the
impurity which brings the estimate $a_0\simeq a$
for neutral impurities.

The asymptotic behavior of \req{eq:scat1} at $|r_1|\gg
\blambda_\varepsilon$ is nothing but the out-going spherical wave,
corresponding to the $s$-scattering. It is important to emphasize
that the dominance of the $s$-channel is the consequence of the
large wave-length and not of the peculiarities of the impurity
potential. The proportionality coefficient between the spherical
wave and the amplitude of the plane-wave is the scattering
amplitude that enables us to find the elastic scattering
cross-section of the electron with the original momentum direction
$\n_i$ to the final direction $\n_f$ \be d\hat{s}=
\frac{d\n_f}{2\pi \lambda_\varepsilon} \left[\hat{\mathbb
P}\left(\n_f\right) \hat{f}_1\hat{\mathbb P}\left(\n_i\right)
\right] \otimes \left[\hat{\mathbb P}\left(\n_f\right)
\hat{f}_1\hat{\mathbb P}\left(\n_i\right) \right]^\dagger.
\label{eq:crosssection} \ee

If there were no interference, \req{eq:crosssection}
would describe all the kinetics of the system.
For the impurity density, $n_{imp}\ll 1/a_0^2$, the mean free path is
estimated from
\be
\ell_{el}\simeq\left({n_{imp}\parallel\hat{ s}\parallel}\right)^{-1}
\simeq {\blambda_\varepsilon}/\left({n_{imp}a^2_0}\right) \gg
{\blambda_\varepsilon},
 .
\label{eq:l} \ee

To understand the role of the multiple scattering, consider the
two impurity scattering, see Fig.~\ref{fig1}a(i)-(ii).

Applying \req{eq:scat1} twice we obtain for $|\R_{12}| \ll
\blambda_\varepsilon \ll |\r_1|$, the outgoing wave
 with $\hat{f}_1\to \delta\hat{f}=\hat{f}_{12}+\hat{f}_{21}$
and
\be
 \hat{f}_{(ij)}=-\hat{f}_i
{\hat{\boldsymbol{\Sigma}}\R_{ij}}
\hat{f}_j
\left({2\pi|\R_{ij}|^2}\right)^{-1},
\ \R_{ij}=\R_i-\R_j,
\label{eq:f12}
\ee for $i,j=1,2,\ i\neq j$, and $a_0\lesssim
|\R_{12}| \lesssim \blambda_\varepsilon$. Equation \rref{eq:f12}
describes the two-impurity scattering amplitude for the given configuration of
the impurities. The
transport, see \reqs{eq:crosssection}-- \rref{eq:l},
is determined by the powers of the   $\hat{f}_{(ij)}$ averaged
with respect to all configurations
\begin{subequations}
\label{2imp} \be
\begin{split}
&\langle\hat{f}_{(ij)}\otimes \hat{f}_{(ij)}^\dagger\rangle=
n_{imp}\int \frac{d^2\R}{4\pi^2R^4}
\left[\hat{f}_i\hat{\boldsymbol{\Sigma}}\R\hat{f}_j \right]\otimes
\left[\hat{f}_j^\dagger\hat{\boldsymbol{\Sigma}}\R\hat{f}_i^\dagger
\right]
\\
& \quad \simeq \left({4\pi}\right)^{-1}
n_{imp}{\mathcal L}
\left[\hat{f}_i\hat{{\Sigma}}_\alpha\hat{f}_j
\right]\otimes
\left[\hat{f}_j^\dagger\hat{{\Sigma}}_\alpha\hat{f}_i^\dagger
\right],
\end{split}
\raisetag{2.3em}{\label{eq:f1212}} \ee where ${\mathcal
L}=\ln\left({\blambda_\varepsilon}/{a}\right)$. Analogously, we
find \be
{4\pi}
 \langle\hat{f}_{(ij)}\otimes
\hat{f}_{(ji)}^\dagger\rangle \simeq -{n_{imp}{\mathcal
L}}
\left[\hat{f}_i\hat{{\Sigma}}_\alpha\hat{f}_j
\right]\otimes
\left[\hat{f}_i^\dagger\hat{{\Sigma}}_\alpha\hat{f}_j^\dagger
\right]. \label{eq:f1221} \ee
Combining \reqs{eq:f1212}--\rref{eq:f1221} we obtain:
\be
4\pi\langle\delta\hat{f}\otimes \delta \hat{f}^\dagger\rangle \simeq
{n_{imp}{\mathcal L}}
 d_{ij}^{i^\prime j^\prime}
\left[\hat{f}_i\hat{{\Sigma}}_\alpha\hat{f}_j \right]\otimes
\left[\hat{f}_{j^\prime}^\dagger\hat{{\Sigma}}_\alpha\hat{f}_
{i^\prime}^\dagger \right], \label{eq:df22} \ee where
non-vanishing coefficients are $d_{12}^{12}=d^{21}_{21}=1$,
$d_{12}^{21}=d_{21}^{12}=-1$, and summation over
all repeating indices here as well as
over $\alpha=x,y$ in \reqs{2imp}, \rref{eq:f121res} is implied.
\end{subequations}

Equation \rref{eq:df22} is the main result of the qualitative
consideration revealing the origin of the logarithmic divergence.
The  Boltzmann kinetic equation systematically neglects those
contributions. SCBA scheme accounts only for the diagonal
components $d_{12}^{12},d^{21}_{21}$ and misses all the other
contributions. For the scalar disorder, e.g. it leads to the
violating the TRS of the problem.

There are two more sources for the $n_{imp}a^2{\mathcal L}$
corrections. One of them is the correlation of the one-impurity
scattering with the two-impurity scattering in which one of the
impurities is visited twice as shown on Fig.~\ref{fig1}a(iii)-(iv). The
corresponding result is easily obtained from the three-impurity
scattering amplitude [cf, \req{eq:f12}]
 \be
 \hat{f}_{(ijk)}=
\left(\hat{f}_i{\hat{\boldsymbol{\Sigma}}\R_{ij}}
\hat{f}_j\hat{\boldsymbol{\Sigma}}\R_{jk}\hat{f}_k \right)
/\left({4\pi^2|\R_{ij}|^2|\R_{jk}|^2}\right) ,
\label{eq:fijk} \ee
 and we find
\be {4\pi}\langle \hat{f}_{(121)} \hat{f}_{2}^\dagger\rangle\simeq
{n_{imp}{\mathcal L}} \left[\hat{f}_1
\hat{{\Sigma}}_\alpha\hat{f}_2\hat{{\Sigma}}_\alpha\hat{f}_1
\right]\otimes \hat{f}_2^\dagger . \label{eq:f121res} \ee
This
correction is missing in SCBA. The last logarithmic
effect arising in this order is the logarithmic dependence
$\propto n_{imp} a^2 \varepsilon\ln\varepsilon$ of the averaged forward
scattering amplitude. It does not affect scattering processes
directly but renormalizes the spectrum of the free Hamiltonian
$H_0$.

Analogously, one considers the scattering from
the three-impurity configurations, see Fig.~\ref{fig1}b), starting from the expression
\rref{eq:fijk} and finds $54$
contributions $\propto {n_{imp}^2{\mathcal L}^2}$ only $6$ of
which are included in the SCBA.

Interference of few processes, is  somewhat
special as they do not vanish completely even if the distance
between impurities is larger than $\blambda_\varepsilon$. In fact the
interference $1-2$ and $3-5$ are the first contributions giving
rise to the WL correction which is also
logarithmic. The WL, however, originates from
the spatial scales larger than $\ell_{el}$, and
that is why they can be separated from the UvLC.

%\newpage

{\em Field theory and RG}--
Let us turn to the  rigorous calculations using the supersymmetry method. Due
to the $4\times 4$ matrix structure of the Hamiltonian, Eqs. (\ref{a1}, \ref%
{a6}), the supervectors $\psi $ should have $4$ times more components than
usually used\cite{book},
i.e we need $32$-components for calculation of the conductivity. The resulting
32d-space can be presented as a direct product of five 2d ones,
 $AB\otimes  KK'\otimes AR \otimes eh \otimes g$, where
$AR$, $eh$ and $g$ are the retarded-advanced, particle-hole, and the fermion-boson sectors.
Averaging
over $\hat{V}$ and using \req{a8}, we find
\bea
&&\langle\dots\rangle=\int \cdots\exp \left( -L\left[ \psi \right] \right){\mathcal D}\psi,
\ \psi^\dagger\hat{\boldsymbol{\Lambda}}=\bar{\psi}=\left[\hat{\mathbb C}\psi\right]^T,
\nonumber\\
&&
\hat{\boldsymbol{\Lambda}}=\hat{\Lambda}\otimes\openone^{AB};\ \
\label{a11}
\hat{\Lambda}=\tau_z^{AR}\otimes \openone^{KK'}\otimes \openone^{eh} \otimes \openone^{g};
\label{psi}\\
&& \hat{\mathbb C}=\hat{C} \otimes \tau_y^{AB}= \hat{z}\otimes
\openone^{AR}\otimes \left(
\tau_-^{eh}\otimes\openone^g-\tau_+^{eh}\otimes\tau_z^g \right).
\nonumber \eea where $\dots$ on the LHS  stand for any combination
of advanced/retarded Green functions $\hat{G}^{A,R}=\left(
\varepsilon\mp\omega/2 -\hat{H}_0-\hat{V} \mp i0 \right) ^{-1}$
and  $\cdots$ in the RHS for the corresponding sources breaking
the supersymmetry. Their form can be found in Ref.~\cite{book} but
it will not be important for us.
%\begin{equation}
%G_{\varepsilon }^{A}\left( \mathbf{r,r}^{\prime }\right) =-2i\int \psi
%_{b}\left( \mathbf{r}\right) \bar{\psi}_{b}\left( \mathbf{r}^{\prime
%}\right) \exp \left( -L\left[ \psi \right] \right) D\psi  \label{a11}
%\end{equation}%
The Lagrangian $L\left[ \psi \right]=
L_{0}\left[ \psi \right] +L_{int}\left[ \psi \right] $ is given by
(sum over $i,m=x,y,z$ is implied)
\be
\label{a12}
\begin{split}
&L_{0}\left[ \psi \right] =i\int \bar{\psi}\left[ \varepsilon -\hat{\mathbb{H}}%
_{0}-\hat{\boldsymbol{\Lambda}}\left(\frac{\omega}{2}+i0\right) \right] \psi d\mathbf{r,}  \\
&L_{int}\left[ \psi \right] =\frac{1}{2}
\int\left[\gamma_0 \left( \bar{
\psi}\psi \right)^2+
\Gamma _{m}^{i} \left( \bar{%
\psi}\hat{\mathbb G}_{m,i}\psi \right) ^{2}\right]d\mathbf{r}
\end{split}
\ee
where $\left(\hat{\mathbb H}_{0}, \hat{\mathbb G}_{m,i}\right)
=\left(\hat{H}_{0}, \hat{G}_{m,i}\right)
\otimes \openone^g \otimes \openone^{AR} \otimes \openone^{eh}
$.

The perturbation theory in $L_{int}\left[ \psi \right] $ leads to
UvLC  and we can calculate the integral (\ref{a11}) using a RG
scheme. We decompose $\psi $ as $\psi =\psi _{0}+\tilde{\psi}$,
where $\tilde{\psi}$ is slow and $\psi _{0}$ is fast, and
integrate out $\psi _{0}$. We rescale $\tilde{\psi}$ and the
coordinates to keep the coefficient in front of $\varepsilon $ and
the ultraviolet cut-off intact. It gives back \req{a12}, with
renormalized couplings $\gamma _{0}, \Gamma_{m}^i$,  Eq.
(\ref{a10}) and the velocity $v$, Eq. (\ref{a1}). This yields RGE
(see Ref.~\cite{ae} for details), which we display  for the most
interesting case $\gamma_0 \gtrsim \Gamma_{m}^i$:
\begin{eqnarray}
&&{2\pi v}\partial_t v =
%-\frac{1}{2\pi v}\left[ \gamma _{+}+2\gamma _{\perp
%}+2\beta _{z}+4\beta _{\perp }\right]
-\left( \gamma _{0}+g_\parallel+2g_\perp\right);
\label{a15}
\\
&&{9 \pi v^2}\partial_t\gamma _{0}\approx
2\left(g_{\parallel}^2+2g_\perp^2\right)
 ;\
{\pi v^2}\partial_t \delta g_{\parallel,\perp}\approx-3\gamma_0\delta g_{\parallel,\perp};
\notag\\
&&{9 \pi v^2}\partial_t g_{\parallel}\approx
-8g_{\parallel}^2-20 g_{\parallel} g_{\perp} +14g_\perp^2;
\notag\\
&&{9 \pi v^2}\partial_t g_{\perp}\approx 4 g_{\parallel} g_{\perp}-18 g_{\perp}^2.
\notag
\end{eqnarray}%
Here $g_{\parallel } =\gamma _{z}+2\gamma _{\perp };$
$g_{\perp } =\beta _{z}+2\beta _{\perp };$
$\delta g_{\parallel }=\gamma _{z}-\gamma _{\perp };$
$ \delta g_{\perp } =\beta _{z}-\beta _{\perp };$
 $t$ is the logarithm of the
running energy.

Solving \reqs{a15} down to the energy, $|\varepsilon|$, we find
\be
{v\left({\varepsilon }\right)}=
\left( \frac{\gamma_0}{\pi}\ln\frac{|\varepsilon|}{\varepsilon_0}\right)^{1/2}\!\!,\ \
\gamma _{0}\left(\varepsilon \right) =\gamma _{0}+{\mathcal O}\left(
\frac{g_\parallel(\varepsilon)}{\gamma_0}
\right),
\label{a16}
\ee
where $\varepsilon_0\simeq J \exp\left(-\pi v_0^2/\gamma_0\right)$ is the energy
at which the first loop RG  breaks down, and $J$ is the bandwidth.

The last of \reqs{a15} yields non-mixed valleys, $g_\perp=0$,
to be unstable, and $g_\perp$ flows towards
\be
\label{a16a}
g_{\parallel }(\varepsilon) \approx g_\perp(\varepsilon) \approx
9\gamma_{0}/
\left\{
14 \ln
\left[  t^{\ast}/
\ln |\varepsilon| /\varepsilon_0
\right]
\right\},
\ee
$\delta g_{\parallel,\perp}(\varepsilon) \propto \ln^3(|\varepsilon|/{\varepsilon_0})$,
and $t^*$ depends on $g_{\parallel,\perp}(\varepsilon\simeq J)$.

Equations \rref{a16} enables one to use the renormalized parameters for the standard calculation
of the transport coefficients. The diffusion coefficient $D(\varepsilon)$ is given by
\begin{equation}
D(\varepsilon)=v^2(\varepsilon)\tau_{tr}(\varepsilon)/2;
\ 1/{\tau _{tr}}(\varepsilon)=\pi \gamma _{0}\nu _{\varepsilon }/4,
\label{a170}
\end{equation}
where $\nu _{\varepsilon }=\left\vert \varepsilon \right\vert/
\left( \pi v_{\varepsilon }^{2}\right)$ is the density of states
(per one physical spin). Einstein relation and \reqs{a16} yield
\cite{prefactor} \be \sigma =
\frac{4e^2}{\pi^2\hbar}\ln\left(\frac{|\varepsilon|}{\varepsilon_0}\right).
\label{a171} \ee Equation \rref{a171} is the universal formula for
the UvLC. It describes either the temperature or the density
dependence of the conductivity limited by the short-range
disordered potential $|\varepsilon|\to {\mathrm
max}\left(\varepsilon_F,T\right)$. It also gives the leading
dependence of the thermopower through the Mott-Cutler formula. At
$|\varepsilon|\lesssim\varepsilon_0$, the logarithms are cut
by $1/\tau_{tr}$, which leads to the replacement of $\ln(\cdot)$ to
the factor of the order of unity. The precise value of this
factor, however, can not be calculated within \reqs{a15}.

{\em $Nl\sigma$-model and localization}-- At distances larger than
$v/\max(|\varepsilon|,\varepsilon_0) $ soft modes giving rise to
UvLC freeze out, and only the degrees of freedom guarded by the
fundamental symmetries of the system remain gapless. All those
degrees of freedom are described by $nl\sigma$-model \cite{book}
and functional integral \rref{psi} is replaced by integral over
$16\times 16$ supermatrices in $KK'\otimes AR \otimes eh \otimes
g$ space \cite{ae}: \be
\begin{split}
&\langle\dots\rangle=\int \cdots\exp \left( -F\left[\hat{Q} \right] \right){\mathcal D}\hat{Q},
\
\hat{Q}^2\!=\openone;\ \hat{Q}=\hat{C}\hat{Q}^T\hat{C}^T\!;
\\
& \hat{Q}^\dagger=\hat{\mathbb K}\hat{Q}\hat{\mathbb K};
\quad \hat{\mathbb K}=
\begin{pmatrix}\openone^{g}& 0\\0 &\tau_3^g& \end{pmatrix}_{AR}\!\!\!
\otimes \openone^{KK'}\!\!\otimes \openone^{eh}.
\end{split}
\raisetag{2em}
\label{Q}
\ee
The free energy functional
$F\left[ Q\right] $ takes the form
\be
\begin{split}
&F =\frac{\pi \nu _{\varepsilon }}{16}Str\int \Bigg\{
D(\varepsilon)
\left( \nnabla \hat{Q}-\frac{ie\boldsymbol{A}}{c} \left[\hat{Q};\hat{\mathcal T}_3\right]\right) ^{2}
+2i\omega \hat{\Lambda} \hat{Q}
\\
&
-\frac{\pi \nu _{\varepsilon }g_\parallel}{4}
\left[ \hat{\rho}_{z},\hat{Q}\right]^2
-\frac{\pi \nu_\varepsilon g_\perp}{4}
\left(\left[ \hat{\rho}_{x},\hat{Q}\right]^2
+
\left[ \hat{\rho}_{y},\hat{Q}\right]^2
\right)
\Bigg\}d\mathbf{r};
\\
&
\left(\hat{\rho}_\alpha,\hat{\mathcal T}_3\right)
=\left(\tau_\alpha^{KK'}\!\!\otimes \openone^{eh},
\openone^{KK'}\!\!\otimes \tau_z^{eh}\right)\otimes\openone^{AR}\otimes\openone^{g},
\end{split}
\raisetag{1.7em}
  \label{a19}
\ee
where $\boldsymbol{A}$ is the vector potential due to the magnetic field normal to
the graphene, and the entries include UvLC.

\begin{figure}[h]
\unitlength=2.3em
{\includegraphics[width=10.8\unitlength]{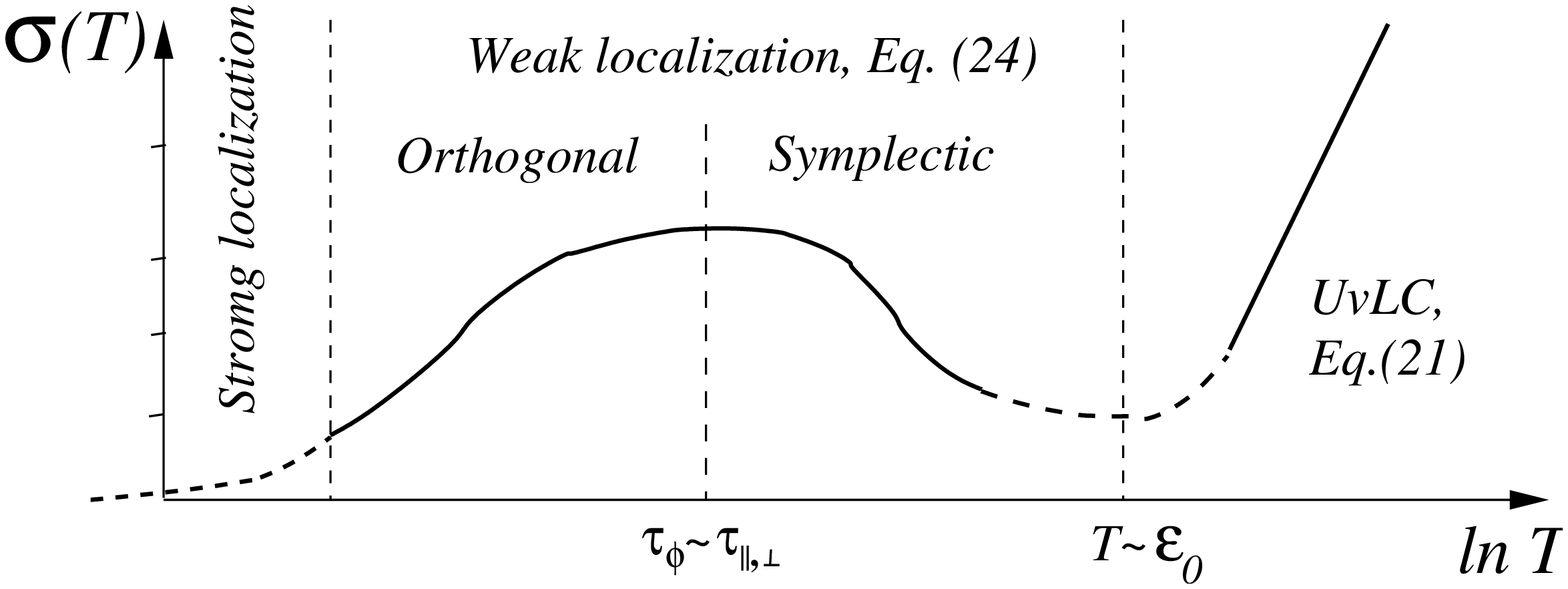}}
\caption{Schematic dependence, $\sigma(T)$,
for the undoped graphene and for $\tau_\phi^{-1}\propto T$.}
\label{fig20}
\end{figure}

Equation \rref{a19} is the only form allowed by the symmetries of
the problem. The symmetries of the $Q$-matrices, \req{Q},
correspond to the two replicas of symplectic ensemble, which would
flow to the limit of large conductances. However, due to
$g_{\parallel,\perp}>0$ only $\hat{Q} \propto \openone^{KK'}$ is
allowed at large distances and one obtains a generic orthogonal
ensemble. Thus, all the eigenstates are localized.
Schematic temperature dependence of $\sigma$ for
undoped graphene is sketched on Fig.~\ref{fig20}, \cite{macdonald}.

The fist loop correction in \req{a19} yields the
WL  \cite{prefactor}
\be \label{a20}
\begin{split}
&\Delta \sigma_{WL}=\frac{e^{2}}{2\pi^2 \hbar}
\sum_{j=0}^{3}d_j
\left[
\ln\left(\frac{1}{-i\omega_j\tau_{tr}}\right)-Y\left(\frac{1}{-i\omega_j\tau_{B}}\right)
\right]
\\
&\omega_0\!=\omega+i0; \ \omega_1\!=\omega+\frac{i}{\tau_\perp};
\ \omega_{2,3}\!=\omega+\frac{i}{2\tau_\parallel}+\frac{i}{2\tau_\perp},
\end{split}
\raisetag{2em} \ee where $-d_0=d_{1,2,3}=1$, $\tau _{\parallel
,\perp }^{-1}=2\pi \nu _{\varepsilon }g_{\parallel ,\perp }$,
$\tau_{B}^{-1}=4D(\varepsilon) eB/(\hbar c)$,
$Y(x)=\psi(x+1/2)+\ln x$ and $\psi(x)$ is the di-gamma function.
Inelastic processes are accounted for by $-i\omega \to
-i\omega+\tau_\phi^{-1}$, where $\tau_\phi$ is the dephasing time.

Equation \rref{a20} agrees with Ref.~\cite{falko}.
The new information here is the logarithmic dependence of the parameters on
the electron energy \cite{warping}, see \reqs{a16}--\rref{a170}.

In conclusion, we have presented a complete description of a disordered
graphene and demonstrated that there are two different types of logarithmic
contributions into physical quantities, see   \reqs{a171} and \rref{a20}.

We are grateful to V. Falko, L. Glazman and P. Kim for remarks
and to Transregio 12 for a support of K.B.E.


\begin{thebibliography}{99}
\bibitem{novo} K.S. Novoselov \textit{et al}, Science \textbf{306}, 666
(2004);
%K.S. Novoselov \textit{et al},
Nature \textbf{438}, 197 (2005);
%K.S.Novoselov \textit{et al, }
Nature Physics \textbf{2}, 177 (2006).

\bibitem{zhang} Y. Zhang \textit{et al, }Phys. Rev. Lett. \textbf{94},
176803 (2005); Y. Zhang \textit{et al, }Nature \textbf{438}, 201 (2005)

\bibitem{berger} C. Berger \textit{et al}, J. Phys. Chem. B \textbf{108}
19912 (2004); Science \textbf{312}, 1191 (2006); J.S. Bunch \textit{et al}, Nano Lett. \textbf{5, }287 (2005);

\bibitem{Pikus}
For a general analysis, see  G.L. Bir and G.E. Pikus,
{\em
Symmetry and strain-induced effects in semiconductors}, New York, Wiley (1974).

%\bibitem{semenoff} G.W. Semenoff, Phys. Rev. Lett. \textbf{53}, 2449 (1984)

\bibitem{gusynin} V.P. Gusynin and S.G. Sharapov, Phys. Rev. Lett. \textbf{%
95, }146801 (2005); D. A. Abanin, P.A. Lee, L.S. Levitov, con-mat/0602645;
J. Tworzydlo {\em et.al.}
%, B. Trauzettel, M. Titov, and C.W.J. Beenakker
;
cond-mat/0603315;  V.V. Cheianov and V.I.
Falko, cond-mat/ 0603624.

\bibitem{Kats}M.I. Katsnelson, cond-mat/0512337; cond-mat/0606611.

\bibitem{shon} N.H. Shon, T. Ando, J. Phys. Soc. Jpn \textbf{67}, 2421 (1998)

\bibitem{suzuura} H. Suzuura, T. Ando, Phys. Rev. Lett. \textbf{89}, 266603
(2002)

\bibitem{zheng} Y. Zheng and T. Ando, Phys. Rev. B \textbf{65},
245420 (2002)

\bibitem{ando} T. Ando, Y. Zheng, and H. Suzuura, J. Phys. Soc.
JPN. \textbf{71}, 1318 (2002)


%\bibitem{mccann} E. MacCann, V.I. Falko, Phys. Rev. B \textbf{71}, 085415
%(2005)

\bibitem{peres} N.M.R. Peres, F. Guinea, and A.H. Castro Neto, Phys. Rev. B
\textbf{73, }125411 (2006)
\bibitem{Morpurgo}A.F. Morpurgo and F. Guinea, cond-mat/0603789.

\bibitem{falko} E. McCann \textit{et al}, cond-mat/0604015



\bibitem{agd} A.A. Abrikosov, L.P. Gorkov, and I.E. Dzyaloshinskii, \textit{%
Methods of Quantum Field Theory in Statistical Physics, }Prentice Hall, New
York (1963)\textit{\ }

\bibitem{book} K.B. Efetov, \textit{Supersymmetry in Disorder and Chaos},
Cambridge University Press, New York (1997).
%; Adv. Phys. \textbf{32}, 53(1983)



\bibitem{footnote1}
The  matrices $\hat{\Sigma}, \hat{z}$ are different from those of
Refs.~\cite{shon,suzuura,peres,falko}, due to the different choice
of the vector in \req{bloch}.

\bibitem{ae} I.L. Aleiner, K.B. Efetov (in preparation).

\bibitem{iordanskii} S.V. Iordanskii and A.E. Koshelev, JETP Lett, {\bf 41},
574 (1985).




\bibitem{morozov} S.V. Morozov \textit{et al, }cond-mat/0603826.

\bibitem{prefactor} Here, we restored $\hbar$ and the spin degeneracy.

\bibitem{warping} Warping terms considered in Ref.~\cite{falko}
can be shown to produce $\tau_\perp/\tau_w \propto \varepsilon^2\ln(|\varepsilon|/\varepsilon_0) \to 0,\ 
\varepsilon\to 0$.

\bibitem{macdonald} Nomura and MacDonald, cond-mat/0606589,
claimed $\sigma\simeq 4e^2/h$ at $\varepsilon \to 0$. We believe this
conclusion to stem from the incorrect form of the disordered Hamiltonian.
\end{thebibliography}
\end{document}